\documentclass[epjH]{svjour}
\usepackage{graphicx}
\usepackage{url}
\usepackage{multirow}
\usepackage{amsmath,amssymb}
\usepackage{soul}

\RequirePackage{color}
\definecolor{MyDarkGreen}{rgb}{0.02,0.60,0.06}

\newcommand{\green}[1]{{\color{MyDarkGreen}{#1}}}

\begin{document}
\title{Schottky's forgotten step to the Ising model}
\author{Reinhard Folk\inst{1}\fnmsep\thanks{\email{r.folk@liwest.at}} 
\and Yurij Holovatch\inst{2,3,4,5}}
\institute{Institute for Theoretical Physics, University Linz, 4040 Linz, Austria \and Institute for Condensed Matter Physics of the National Acad. Sci. of Ukraine, 
79011 Lviv, Ukraine \and ${\mathbb L}^4$ Collaboration \& Doctoral College for the
Statistical Physics of Complex Systems,
Leipzig-Lorraine-Lviv-Coventry, Europe \and
 Centre for Fluid and Complex Systems, Coventry University, Coventry
CV1 5FB, UK \and
Complexity Science Hub Vienna, 1080 Vienna, Austria
}
\abstract{A longstanding problem in natural science and later in physics was the understanding of the existence of ferromagnetism and 
its disappearance under heating to high temperatures. Although a qualitative description was possible by the Curie - Weiss theory it 
was obvious that a microscopic model was necessary to explain the tendency of the elementary magnetons to prefer parallel ordering at low temperatures.
Such a model was proposed in 1922 by W. Schottky within the old Bohr - Sommerfeld quantum mechanics and claimed to explain the  high values 
of the Curie temperatures of certain ferromagnets. Based on this idea Ising formulated a new model for ferromagnetism in solids. Simultaneously the old quantum mechanics was replaced by new concepts of Heisenberg and Schr\"odinger and the discovery of  spin. Thus Schottky's idea was outperformed and finally replaced 
{{in}} 1928 by Heisenberg exchange interaction.
This led to a reformulation of Ising's model by Pauli at the Solvay conference in 1930.
Nevertheless one might consider Schottky's idea as a forerunner of this development explaining 
and asserting that the main point is the Coulomb energy leading to the essential interaction of neighboring elementary magnets.
} 
\maketitle

\section{Introduction}\label{I}
\label{intro}
Simple models often turn out to survive for generations and become cornerstones of teaching and research  in physics. One of such models is 
the Ising model\footnote{As Barry McCoy notes in his lectures \cite{McCoy95}: {\it ``The Ising model has led to developments in mathematics 
which have been widely applied to the theory of random matrices, to quantum gravity, and the Ising correlations themselves are
directly related to N = 2 supersymmetric quantum field theory in two dimensions.''}}, another one is Landau's model,\footnote{Michael E. Fisher 
called Lev Landau  {\it ``the founder of systematic effective field theories, even though he might not have put it that way.''}
(\cite{Fisher99}, p. 91)} where Landau introduced the concept of the order parameter (\cite{landau}).
The Ising model was first developed for understanding the phenomenon of magnetism known for centuries. It transformed over the time into an  
intellectual tool in other fields of science. Very recent outlying examples might be found in social sciences \cite{Hurtado-Marín21},
ecology \cite{Noble18},  linguistic\green{s} \cite{DeGiuli19}.

The idea for Ising's model started from searching concepts to explain  the emerging property of ferromagnetism in a solid state system. 
But the underlying microscopic concepts (the quantum mechanics) were  more and more in question so that its macroscopic mathematical 
formulation (by statistical physics) changed when new microscopic concepts were presented. Surprisingly, {{as we show in this paper,}} these conceptual changes
and formulations were to a large extent done by scientists working at the same university or even the same institute without 
recognizing that their ideas  come together at the same goal. 

A side effect of the revolutionary changes in quantum mechanics was, that old concepts were lost in the presentation of new ideas. This we consider not as a question of negligence but as a sign of the deep changes in physical thinking made by the new concepts.
One of such examples is Schottky's suggestion to explain the ferromagnetic ordering of the elementary magnetic units in a solid by 
a certain arrangement of the electrons in atoms in order to minimize their Coulomb energy. It is the intention of this paper to 
show how this idea was replaced by Pauli's new binary quantum number for the electron and Heisenberg's exchange force and how it 
led to the reformulation of Ising's model by Pauli. By no means we want to rise by this analysis
a question of priorities, our goal is rather to show that the Ising model has a very rich history to which also the work by Schottky belongs.

The rest of the paper is organized as follows. In the next section we briefly 
analyse the historical situation in physics of magnetism at the beginning of 1920. Lenz idea about the elementary
magnet flipping process as a ground for macroscopic magnetism is discussed in section \ref{III}  together with the Stern-Gerlach experiment. In section \ref{V} we discuss a model suggested by Schottky during his work
in Hamburg university (the same place, where Lenz, Stern, Gerlach and Pauli conducted their research and where Ernst Ising
later defended his doctoral thesis). Schottky's model was based on the idea about synchronization between the electrons in a 
molecule or in a solid \cite{schottky}. 
{{Section \ref{VI} describes how Ising used Lenz' and Schottky's ideas to formulate a new model for explaining ferromagnetism without an explicit Hamiltonian. The changes of the microscopic basis of Ising's model - the new understanding of quantum mechanics, the discovery of the spin and the exchange interaction -  are explained in section \ref{VII}. Section \ref{VIII} finalizes the development with the presentation of a Hamiltonian for Ising's model  after Dirac had found a complete description of the electron including its spin.}}
{{We end by a concluding remark in section \ref{IX} on the importance of the new presentation of Ising's model for it's solution.}} In the Appendices we provide a timetable for some events discussed in our paper (appendix \ref{timetable}) as
well as give explicit calculations  of the synchronisation phenomenon (appendix \ref{calculation}) along the lines suggested by Walter H. Schottky.

\section{Situation after the Great War}\label{II}
At the beginning of the 1920 both microscopic (that is atomic) magnetism and macroscopic (that is ferro-) magnetism posed to 
physicists great puzzles. Whereas the first problem belonged  to the field of quantum mechanics the second one belonged to the field of 
statistical physics.\footnote{{\it It was evident...that the magnetization of an iron bar is modified by temperature variations or mechanical 
constraint and thus does not depend solely on the characteristics of an atom isolated from its neighbors. It was therefore necessary to 
approach the observed macroscopic phenomena by way of interactions involving more than one atom or molecule.(\cite{hoddeson} p. 23)}}

Moreover only two fundamental forces were known at that time - gravitation and the electromagnetic force. A third force - the strong force 
(relevant in the nucleus of an atom) - was postulated in the 1920 but found later in the 1970 and the weak force (also relevant in 
the sub-atomic region) was found in 1934. Thus only the electrostatic or magnetostatic force could be involved in the explanation of ferromagnetism. 
However quantum mechanical constraints on the otherwise classical microscopic calculations might play an important role.

Niels Bohr already in 1911 and Hendrika Johanna van Leeuwen later in 1919  derived in their PhD theses the famous theorem for classical 
nonrelativistic electrons using Maxwell's equations and statistical mechanics stating that: 
{\em at any finite temperature, and in all finite applied electrical or thermal fields, the net magnetization of a collection of electrons 
in thermal equilibrium vanishes identically.} 
However it is not known how widespread this theorem was present in the scientific community.\footnote{Elliot wrote in his historical 
overview of the {\em Developments in magnetism since the second world war} \cite{elliott}, when he commented the Conference on Magnetism in 
Strasbourg 1939: {\it ``One of the most bizarre of these was Miss van Leeuwen’s theorem (van Leeuwen, 1921) which demonstrated that in classical statistical 
mechanics the magnetic susceptibility must be zero. (Bohr in his 1911 dissertation had already gone some way towards a similar result). The reason why 
Langevin’s formula violated this theorem, while giving the physically correct result, lay in his assumptions about fixed magnetic dipole moments which 
were at variance with the strictly classical conditions."}} Anyway the importance of this theorem was  recognized and prominently presented in the year 
1932 by van Vleck (see chapters 24 - 27 in \cite{vanVleck}). 

Thus one might have concluded: when classical Boltzmann statistics is applied  to any dynamical system, the magnetic susceptibility is zero 
(p. 95 \cite{vanVleck}).  Even today one finds the opinion {{that }} {\it a starker conflict between theory and experiment would be hard to imagine: classical physics 
gives no ferromagnetism, no paramagnetism, no diamagnetism, in fact no magnetism 
at all!}\footnote{Department of Physics Trinity College Dublin \url{https://www.tcd.ie/Physics/research/groups/magnetism/facts/guide/understanding.php}}
But this holds not as strict as it was formulated. Although this theorem is very general and its proof is rigorous it does 
not eliminate all possibilities of using classical physics. Indeed {\it ``classical electrons cannot move in a circular orbit around the atomic nucleus...
But many of today's 'classical' theories only use the result of quantum mechanics to force the electron into such orbits, and calculates its radius classically''} (\cite{aharony} p.7). Thus one might use {\it adjusted}  classical concepts in order to attack physical problems as it was done in the old quantum mechanics with the concept of Bohr's magneton. Even after the progress made by the new quantum mechanics and the discovery of the spin the classical pictures remained (see Fig. \ref{bozoroth} at the end of Sec. \ref{VIII}). It was Walter
Schottky, who without knowing about the 
spin degree of freedom developed in 1922 a microscopic model for ferromagnetism in this sense.

In the period we consider here, many problems were related to the influence of a 
magnetic field on atoms and molecules. One of them was the understanding of the splitting 
of spectral lines in magnetic fields. The Bohr-Sommerfeld quantum mechanics coupled the external magnetic field to the electron circling planetary 
like around the nucleus, thereby creating a magnetic moment connected to the angular momentum of the electron.
Due to quantization conditions of the three `action'-variables in the three dimensional mechanics a controversial discussion took place about the 
behavior of the magnetic moment of the atom within the magnetic field known under the term spacial or directional quantization. 
In general it was a common idea to allow for appropriate degrees of freedom of a classical problem only discrete values\footnote{For a very short 
description see chapter 8/\S 10 in Karl Herzfeld's contribution to M\"uller-Pouillets {\it Lehrbuch der Physik} \cite{herzfeldMP}.} - in total three - 
corresponding to the  three degrees of freedom for a particle in three dimensional space.

\section{Lenz's idea {{and Stern-Gerlach experiment}}}\label{III}

Wilhelm Lenz criticized \cite{lenz} the previous theories to explain the Curie law  for the assumption of free rotatability of 
elementary magnets in solids with  reference to Born and his concept of crystal symmetry. Instead he assumed that there are certain 
equivalent directions given by the crystal structure for the elementary magnets in which the latter can be oriented. { That means certain 
direction of the planes in which the electrons circulate.} In the simplest case (e.g. cubic crystals) only one direction and its reversed 
are equivalent. Thus instead of free rotation of elementary magnets he allowed only flipping processes. In this way he could derive Curies 
law by using Boltzmann's statistical theory. 

In order to explain ferromagnetism he speculated that there is an energy difference between the two possible directions of the elementary 
magnets and that this energy is not of magnetic nature. However its origin had to be considered as completely unknown. 

{{At the time when Lenz published these ideas, he was an}}
 extraordinarius professor at the University of Rostock. He applied and succeeded in 
1921 for the chair of Theoretical Physics of the University of Hamburg.  
 {{The same year Stern}} got the position of Lenz in Rostock {{and}}  suggested  an experiment to solve the problem of the spacial or directional quantization. {{The experiment}} was carried out by {{Stern}}  and Gerlach in Frankfurt one year later.  {{They}} wanted to disprove  the Bohr-Sommerfeld model of the atomic orbits but instead it seemed to confirm the picture. The classical degrees of the circular momentum would lead to three values of the quantum number $m=1,0,-1$ in the direction of the external magnetic field. Although the experiment showed only two lines  no one raised  questions, not even Pauli nor Heisenberg. For a historical description of this experiment see \cite{huber,sauer}, for a detailed theoretical analysis see \cite{wennerstroem}. In fact as we know today it was the first experiment to prove the existence of the spin of the electron. Anyway it gave a basis to Lenz idea of distinct direction in solid systems for the elementary magnets in solids contrary of  free rotation (for a more elaborated discussion see \cite{Niss}). In 1923 Stern became director of the newly founded Institut für Physikalische Chemie at the University of Hamburg, where he  was in close contact with Pauli.

\section{Schottky's idea}\label{V}

On 19 September 1922, Schottky began to work as a salaried assistant at the Physical State Laboratory in Hamburg, 
on leave from W\"urzburg (where he habilitated) for the whole winter semester of 1922/1923.\footnote{See e.g.  Encyclopedia.com
the web pages \url{https://www.encyclopedia.com/science/dictionaries-thesauruses-pictures-and-press-releases/schottky-friedrich-hermann} 
and Catalogus Professorum Rostochiensium [2004-]  \url{http://cpr.uni-rostock.de/resolve/id/cpr_person_00002340} In most of the biographies this employment in Hamburg is not noted. On the motivation for Schottky to go to Hamburg one may only speculate since there are no further documents available (see p. 222 \cite{serchinger}).}
Schottky was already financially independent at that time, as his patents (from the years 1916 - 1919 at Siemens AG) brought in higher royalties than his salary \cite{endres}.
It was in this time when he formulated his model {\em About 
the Rotation of the Atomic Axes in Solids (\"Uber die Drehung der Atomachsen in festen K\"orpern)} \cite{schottky} and one can assume that he also was in discussion with Lenz.  {{Indeed}} Schottky 
thanks Lenz in his publication \cite{schottky} for bringing some papers to his attention, also Pauli was present when Schottky {{reported}} 
the paper at the Verhandlungen der Gesellschaft Deutscher Naturforscher und \"Arzte (Meeting of the Society of German Natural Scientists and 
Physicians) on September 21 1922 in Leipzig. Although Heisenberg was not present he asked Schottky for more information in a letter from October 7 and Schottky sent him his publication \cite{serchinger}.

The model Schottky suggested is based on the idea that between the electrons in a molecule or in a solid a {\em synchronization} takes place. 
This synchronization then is responsible for a force between the atom forming the molecule or solid. It was developed at that time for 
the problem to calculate systems with more than one electron as in molecules or solid state systems. Nowadays one would say the problem 
is to calculate the many-body wave function but at that time it was unclear how to use the quantum mechanical conditions for these more complicated 
classical systems. {Max Born \cite{born22} described this procedure as an onset of a  phase relation  between 
two electrons when two one-electron atoms approach each other and begin to interact. This can be compared  to 
the synchronization of two coupled classical oscillators. Later the concept of synchronism was also applied 
to the case where two electrons are circling around one nucleus using classical analysis of motion of two 
planets around the central star.}
Albert Land\'e called it lattice-synchronism (Gittersynchronismus) leading to the cohesion force in diamonds \cite{lande}. He argued that in this way (by the phase relation) a regular lattice symmetry is established. This in fact can be considered as an example for creating a long-range order out of a short-range interaction.

{Not only the phase is of importance in the synchronization processes discussed above,  but also the direction of rotation of 
the electron on its orbit around the atomic nucleus.} 
Land\'e chooses  two electrons rotating {in opposite directions} since then the resulting 
elementary dipoles attract each other. 
He thanks Wilhelm Lenz in a footnote for drawing  his attention to this fact.  Lenz had already in 1919 contributed results on H$_2$ molecules 
\cite{lenzH2} but did not publish further calculations. The topic was taken up shortly later by Lothar Nordheim in his thesis \cite{nordheim} 
under his supervisor Max Born in G\"ottingen. Thus {\em synchronism} was a concept used in the quantum mechanical treatments of many-electron systems.

\subsection{The effect of synchronization on the ordering in ferromagnets}

\begin{figure}\centering
\includegraphics[height=5.5cm]{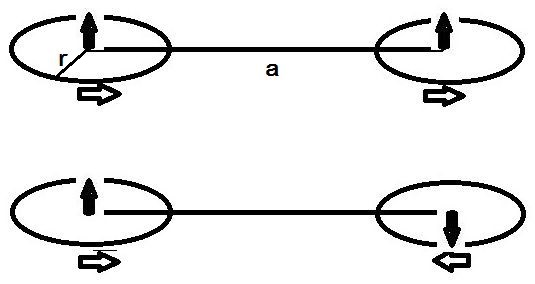}
\caption{Two electrons circling in the same or opposite direction  above or beside each other. Schottky compared the energy difference between the left and right resp. the upper and lower configuration of the two electrons, see Eq. (\ref{schottkyresult}). \label{config}}
\end{figure}

Schottky applied the idea of synchronism combined with the direction of rotation of the electron around the atomic nucleus to the problem of 
magnetic ordering. The problem was to find out how a two electron system, especially its interaction is described and how the angular 
momenta, respectively the resulting elementary magnetons, are arranged. Assume two atoms (at distance $a$) each with one electron circling 
(radius $r$) around the nucleus in planes {\em parallel} to each other with the same frequency $\omega$, as shown in Fig. \ref{config}. 
The parallelism of the orbitals is important since then the resulting elementary magnetic moments point along (or against) the same direction. 
The electrostatic energy $E$ of the two electrons should be as small as possible during the circling around the nucleus. This is achieved 
by keeping the distance between the two electrons during the circling as large as possible. Since the distance and the radius of the 
orbit is fixed it is the phase between the two circling electrons which is adjusted in such a way that the Coulomb energy is as small as possible.

Then the direction of one of the electrons is reversed, so that the induced magnetic moments are antiparallel. Again the electrostatic energy 
is calculated. The difference between these two electrostatic energies defines if parallel or antiparallel magnetic moments are preferred. 
The difference in the energy due to the magnetic dipole interaction might be neglected if it is much smaller then the electrostatic energy difference.

This may be considered a synchronization effect as it is known from mechanics for coupled pendulums since Huyghens. Nevertheless in quantum 
mechanics this mechanism remained obscure. Indeed as it turned out later it is the mechanism of the Pauli principle which influences the charge 
distribution of the two electrons.

In his publication \cite{schottky}, Schottky considered two configurations of the circling electrons in a solid (e.g. of cubic structure) above or beside each 
other, see Fig. \ref{config}. In the first case the elementary magnets are parallel, in the second case they are perpendicular to the connecting 
axis between the centers of rotation. In Fig. 1 of his paper Schottky showed only one configuration  but discussed both cases. 

The centers of the circles are within  the distance $a$ typically of a lattice distance, the radius $r$ of each circle is of atomic size. One has 
to calculate  the distance of the two electrons during the circulation with frequency $\omega$ when they are placed on the circle at the 
beginning  with a certain phase $\phi$. The phase with the smallest Coulomb energy
is taken for this case. Then the direction of one circling electron is reversed and a new phase is searched for which the Coulomb energy 
is calculated. Schottky gives (without explicite calculations) a result of his consideration
\begin{equation}
\Delta E= \frac{e^2}{a}\Big(1 - \frac{1}{\sqrt{1+k(\frac{r}{a})^2}}\Big) \, . \label{schottkyresult}
\end{equation}
The value of $k$ depends on the configuration. If the electrons are above each other $k=4$, if they are besides each other it is 
{\it ``about'' (``etwa gleich")} $2$. Inserting the usual magnitudes for the atomic distances ($r/a$ from $1/4$ to $1/2$) lead to an energy difference whose corresponding characteristic temperature, $T\approx \Delta E/k_B$ (where $k_B$ is the Boltzmann constant) is of the order the experimental Curie temperatures. The magnetic dipole interaction of elementary dipoles would only lead to characteristic temperature of the order of one degree as Schottky remarks. The interaction depends  whether  the connection of the two centers is perpendicular to the planes of  the orbits or lies 
within the plane of two orbits. In consequence in a cubic spacial model the interaction is anisotropic. In  Appendix \ref{calculation} we give a more detailed calculation. 
Schottky points out that an essential condition for this result is the {\em absolutely strict synchronism} between neighboring circling electrons 
otherwise no energy difference would be present if the circulation direction is changed. He also remarks that this feature does not play any role 
for the magnetic interaction of the atoms (i.e. their dipole moments). Indeed the dipole interaction is so small that the energy difference between 
parallel and antiparallel dipoles is less than one degree. This weak magnetic interaction force was already a problem in understanding ferromagnetism
by the Curie-Weiss theory. It was unclear how the necessary strength of the assumed internal field ({similar} to the internal pressure of the Van 
der Waals theory) could be generated.

\subsection{A more detailed calculation}

Since it is unclear how Schottky arrived at his results a more detailed calculation is presented in Appendix \ref{calculation}. This leads to a  somewhat different estimate but his argumentation works in all cases where the circling takes place either in one plane or in parallel planes. Just averaging the distance of the electrons over one circulation and calculating the difference in the Coulomb energy one obtains 
\begin{equation}
\Delta E= \frac{e^2}{a}\Big(\frac{1}{\sqrt{1+2(\frac{r}{a})^2}} - \frac{1}{\sqrt{1+4(\frac{r}{a})^2}}\Big)\, .  \label{ourresult}
\end{equation}
This leads to a smaller difference compared to Schottky's result but it remains within the wanted order for the Curie temperature. The advantage is that it 
only depends on the distance {{between}} the two electrons and not on the orientation in space. Thus the effect is the same wether the atoms are above or beside each other.

\section{After Schottky's paper} \label{VI}

After his time at Hamburg university Schottky got {{in}} 1923 the chair of Theoretical Physics at  the University of Rostock, whereas Otto Stern  moved to the University of Hamburg. In 1927 then Schottky finally returned to Siemens \cite{welker}.

There are almost no citation of Schottky's publication in connection with his idea about the interaction of the elementary magnetons in the field of magnetism. 
Nevertheless the paper became famous {{because of what is now termed as}}  {\em Schottky anomaly}. However 
in most of the cases nowadays no citation of the original paper is given \cite{westrum}. Schottky calculated in his paper the specific heat of a two level system 
(the energy levels are separated by the above mentioned energy difference) showing a peak in the low temperature region. The observation of such a peak 
in the specific heat is usually taken as {{an}} indication of the existence and the spacing of energy levels in a system.

It seems to be reasonable that Schottky's paper was the incentive for Lenz to suggest Ising as a topic for his thesis to formulate this idea in a statistical 
model showing the ferromagnetic phase (for Ising's life and the model see  \cite{tising} and references therein). Lenz had already suggested Thomas Schr\"oder 
an electrodynamical problem for the thesis, which was published in
1922 \cite{schroeder22}.  Ernst Ising came to Hamburg University in 1921 when Lenz was still in Rostock, he became interested in theoretical physics after Lenz 
got the chair. He joined his group and at the end of 1922 he started to work on his thesis on ferromagnetism. Almost at the same time Lucy Mensing\footnote{A 
lifelong friendship resulted from this time (see Ref. \cite{muenster} for  Mensing's fate).} worked on her thesis, which studied the broadening of spectral lines 
in electric fields using the old Bohr Sommerfeld quantum mechanics. She was mainly supervised by Pauli and her result was published shortly after Ising's paper 
\cite{mensing24}. 

In a letter to Brush Ising described his time in Hamburg and the scientific discussion he had \cite{brush}: {\it ``At the time I wrote my doctor thesis Stern and 
Gerlach were working in the same institute on their famous experiment on space quantization. The ideas we had at that time were that atoms or molecules of 
magnets had magnetic dipoles and that these
dipoles had a limited number of orientations. We assumed that the field of these dipoles would die down fast enough so that only interactions should be taken 
in account, at least in the first order [. . . ]. I discussed the result of my paper widely with Professor Lenz and with Dr. Wolfgang Pauli, who at that 
time was teaching in Hamburg. There was some disappointment {that} the linear model did not show the expected ferromagnetic properties."}

\subsection{Ising's thesis}

Lenz had written \cite{lenz} in his conclusion:\footnote{\it Die magnetischen Eigenschaften der Ferromagnetika werden dadurch auf nicht magnetische Kr\"afte 
zur\"uckgef\"uhrt in \"Ubereinstimmung mit der Aufassung von Weiss, der durch Rechnung und Versuch \"uberzeugend dargetan , da{\ss} das von ihm eingef\"uhrte 
und die Verh\"altnisse in gro{\ss}en Z\"ugen gut darstellende Eigenfeld nicht magnetischer Natur ist. Es ist zu hoffen, da{\ss} es gelingt, auf dem angedeuteten 
Weg die Eigenschaften der Ferromagnetika zu erkl\"aren} {\it``The magnetic properties of ferromagnetics would then be explained in terms of nonmagnetic forces, 
in agreement with the viewpoint of Weiss, who has by calculation and experiment established that the internal field, which he introduced and which generally 
gives a good representation of the situation, is of a nonmagnetic nature. It is to be hoped that one can succeed in explaining the properties of ferromagnetics 
in the manner indicated."} (translation by Brush in his review \cite{brush}).

This was the task Ising had to do. In his thesis he attacked the problem assuming Schottky's idea as microscopic basis: {\it ``Apart from an applied external magnetic field, the elements should be influenced by the forces that they exert on each other. We can give no further details about these forces, which may be of an electrical nature (see W. Schottky, Phys. Zeitschr., 23, 448, 1922); however, we assume that they rapidly reduce with distance, so that in general, as a first approximation, we need only consider the effect of neighboring elements. The latter assumption is to some extent in contrast with the hypothesis of the molecular field, which P. Weiss (C.R. 157. 1405. 1913. and C.R. 158. 29. 1914.) has shown cannot be magnetic in nature.''} (from \cite{thesis}, p. 4).
And then he started the macroscopic calculation following the program known from statistical mechanics developed by Boltzmann and Gibbs. One may note that (1) the circulation of the electrons is taken in fact as an 'order parameter' and (2) the energy difference of the two configurations is taken from Schottky's microscopic model. Moreover, although the treatment follows the classical statistical method, the model is already quantum mechanical due to the spatial quantization and the assumption of  {an elementary magnetic unit} (Bohr's magneton).
 He tried to calculate the partition function of a three dimensional 
system of atoms with elementary magnetons within an external magnetic field in order to find the corresponding free energy as function of the magnetic field 
\cite{thesis}.\footnote{See:
http://www.icmps.lviv.ua/ising/books.html An excerpt
of the thesis ``Contribution to the Theory of Ferromagnetism"
translated by Jane Ising and Tom Cummings can be found on the webpage of the Bibliotheca Augustina. 
A complete translation by B. Berche, R. Kenna and the authors is in preparation.} 
From the derivative with respect to the external field then he would receive the desired magnetization as a function of  external field and temperature. Setting 
the external field to zero would show if a finite magnetization remained in the low temperature region.  

\begin{figure}[h]\centering
 \includegraphics[height=0.12\textwidth]{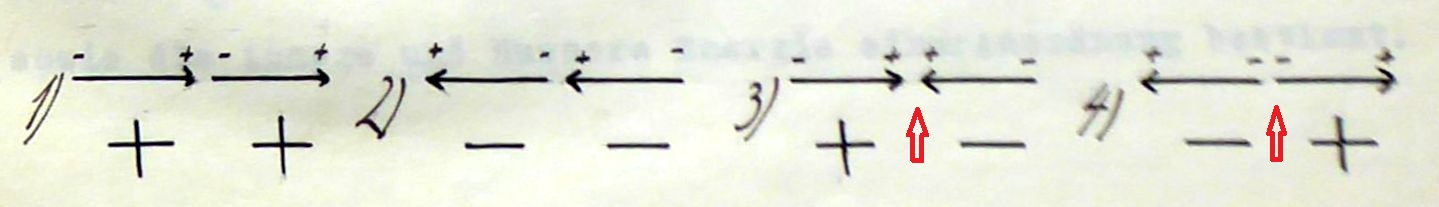}
\caption{Ising's representation of the ordering: Antiparallel configurations cost energy (red arrows), the direction of the arrows are irrelevant and therefore 
Ising replaced them by + and -  \label{ordering}. The picture is taken from the original Ising's thesis, red arrows are added by us.}
 \end{figure}
 
In order to calculate the partition function he had to find the Boltzmann weights  of a configuration of the microstates. There are only two different energies: 
one for parallel and one for antiparallel elementary magnetons (see Fig. \ref{ordering}).  At this point according to citation of Schottky's paper it is clear the energy difference between the two configurations is of the right order to lead possibly to a phase transition with a realistic Curie temperature. No other suggestion for a non-magnetic source of this energy difference was published in literature even for the next years. 
The direction of the arrows is not important. Nowadays we understand 
that the order parameter is not a vector but a scalar quantity, which intuitively has been taken into account by Ising representing the configurations just by 
pluses or minuses \cite{thesis}. Of course the orientation of the elementary magnetons in an external magnetic field depends on its direction. {Due to the method of calculating the magnetization it was essential to include an external field. It was taken along the axis of the 
geometrical arrangement, e.g. along the chain. }

The situation becomes more complicated when Ising tries to extend his model to higher dimensions. In the extension to a planar arrangement  Ising 
commented:\footnote{\it ``Es ist zweckm\"assig, bei diesem in Figur 8 angedeuteten Modell die Anordnung der Elemente in anderer weise aufzufassen; man kann n\"amlich 
sagen, die Elemente sind in $n_1$ Querreihen angeordnet, von denen jede $n$ nebeneinander liegende Elemente enth\"alt. Von einer solchen Querreihe gilt aber ganz 
abgesehen von dem gar nicht in Betracht kommenden Umstand, dass die Richtung des \"ausseren Feldes und der Dipolmomente jetzt normal zur Ausgangsrichtung der 
Elemente liegen, genau dasselbe, was wir oben \"uber die einfache Kette gesagt haben. Daraus folgt ... das mittlere Moment unseres Modells"} {\it ``In this model, 
which is shown in Fig. 8 {[of the thesis; showing a part of elementary magnetons in a plane]}, it is expedient to consider the arrangement of the elements in 
a different manner; one can say that the elements are arranged in $n_1$ transverse rows, each of which contains $n$ {{adjacent}} elements. Of such a transverse 
series, however, apart from the fact that the direction of the external field and of the dipole moments is now normal to the arrangement direction of the elements, 
is exactly what we have said above about the simple chain. From this follows the mean moment of our model."} (from  \cite{thesis} p. 24)
Thus the {\em orbits} of the electrons are above each other {\em and} besides each other and of course perpendicular to the plane. {Therefore the model according 
to Schottky's calculation is now {\em anisotropic}}, see also the remarks of Ising to the spatial model in his publication. If the elementary moments would be perpendicular to the plane, the orbits would be only beside each other. 
The direction of the elementary magnets  in the plane is defined by the direction of the external magnetic field and the orbits are always perpendicular to the 
direction of the external field. 

Niss commented \cite{Niss}: {\it ``Schottky and Ising had different views on the orientation of the elementary magnets: Schottky took them to be pointing perpendicular to the plane in which they lay, while Ising always depicted them as pointing in this plane, at least in the case of a linear chain of them, which was his main object of study. Thus, if the chain extends horizontally, it consists of elementary magnets pointing left or right,
 $\leftarrow$ and $\rightarrow$, respectively (and thus not the way the Lenz-Ising model is usually presented in modern textbooks). I conclude that because Schottky’s argument does not apply to Ising’s linear chain, it very likely did not stimulate Ising’s conception of it."} The elementary magnets are always perpendicular to the orbits of the electrons, therefore it is the plane of the orbits which defines the orientation with respect to a chain axis.  Schottky considered all orientations as it is necessary for solids e.g. of cubic structure. Ising took this in consideration when he treated the double chain (see Fig. 9 in the thesis \cite{thesis}). But another important point to rely on Schottky's idea was that the energy difference of the configurations of the elementary magnetons resulted from the electrostatic force.
 
We know now {that the calculation Ising wanted to perform}  was a mission impossible not only at that time. Even now an exact calculation of the partition function 
{\em with an external magnetic field} is not possible in two and three dimensions. But he solved instead the one dimensional model (the Ising chain) and came to 
the correct conclusion that no magnetic phase exists at a finite temperature. He concluded:\footnote{\it ``... so gelangen wir bei unseren Annahmen nicht zu einer 
Erkl\"arung des Ferromagnetismus. Es ist zu vermuten, dass diese Aussage auch f\"ur ein r\"aumliches Modell zu trifft..." \cite{thesis}} 
 {\it ``... thus with our assumptions we do not arrive at an explanation of ferromagnetism. It is to {\em suppose} that this statement also applies to a spacial 
 model,..."} \cite{thesis}.  

Ising asked for promotion on July 8 1924 and got his degree on
November 3 1924 \cite{reich}. In his judgment of Ising's thesis\footnote{Document UA HH 364 – 13 
Fakult\"aten/Fachbereiche der Universit\"at, Mat.Nat.Prom. 135. 
R.F. thanks Karin Reich for sending a copy of this document.
} Lenz declared the failure of this idea, 
he wrote:\footnote{\it ``Eine befriedigende Theorie mu{\ss} begr\"undet werden auf das Verhalten der Atome eines 
festen K\"orper, wozu die Bohrsche Theorie die erforderlichen Anhaltspunkte liefert. Im Anschluss an eine dieser 
Anforderungen gen\"ugende Theorie des thermischen Verhaltens der paramagnetischen Salze habe ich dem Kandidaten 
vorgeschlagen, diese Vorstellungen auf den Ferromagnetismus auszudehnen. Die ziemlich verwickelten Wahrscheinlichkeitsbetrachtungen 
sind vom Verfasser mit beachtenswertem Geschick durchgef\"uhrt worden. Sie f\"uhrten zum Ergebnis, dass auf dem eingeschlagenen 
Weg ein Ferromagnetismus nicht zustande kommt. Die Gr\"unde hierzu werden diskutiert. Da eine Ab\"anderung der Grundvorstellungen 
\"uber die Atomeigenschaften und ihre Wechselwirkung nicht wohl in Frage kommt so entsteht die Frage, ob der ferromagnetische 
Zustand \"uberhaupt als thermischer Gleichgewichtszustand betrachtet werden kann. Die Untersuchung dieser M\"oglichkeit w\"urde 
indessen den Rahmen dieser Doktorarbeit weit \"uberstiegen haben."} 
{\it ``A satisfactory theory must be based on the behavior of the atoms of a solid, where Bohr's theory provides the necessary clues. 
Based on a theory of the thermal behavior of paramagnetic salts  that met these requirements, I suggested that the candidate 
extends these ideas to ferromagnetism. The rather intricate considerations of probability have been carried out by the author 
with remarkable skill. They led to the result that ferromagnetism does not occur on the path taken. The reasons for this are 
discussed. Since a change in the basic ideas about the atomic properties and their interaction is out of the question, 
the question arises whether the ferromagnetic state can be considered as a thermal equilibrium state at all. Investigating 
this possibility would, however, have far exceeded the scope of this doctoral thesis."}

After finishing his thesis Ising prepared a publication which was received on December 9 1924 and published in Zeitschrift 
f\"ur Physik \cite{Ising25} at the beginning of the year 1925. In the publication he tightened his conclusion to:\footnote{\it ``Es wird gezeigt, 
da{\ss} ein solches Modell [das eindimensionale] noch keine ferromagnetischen Eigenschaften besitzt und diese Aussage wird auch auf 
das dreidimensional Modell ausgedehnt."} {\it ``It is shown that such a model [the one-dimensional] does not yet have ferromagnetic properties 
and that this statement {\em also extends} to the three-dimensional model."}

Although Ising {\em tried to extend} the model in his thesis to higher dimensions but of course did not succeed. All his 
extensions contained an external field and cannot be solved in the way he had in hand at that time. Moreover in the publication  
Schottky's argument is completely eliminated and reduced to just an assumption to hold for the model. No citation of Schottky's paper is included.

Ising left  Lenz group at the University Hamburg in 1925 and began to work in Berlin  at AEG (General Electric Company). His further fate decoupled him from the ongoing research in the field of ferromagnetism and only after the end of World War II he became aware of the importance of the model he had treated in his thesis \cite{tising}.

\subsection{Reaction to Ising's paper}

Karl Herzfeld was the first to refer  \cite{herzfeld} to Ising's publication in the year of its publication.  He also refered to Schottky's idea, in fact this is the only citation in this context which could be found. Herzfeld\footnote{Herzfeld studied in Vienna and his 
thesis was supervised by Hasen\"ohrl. After five years in Munich he became in 1925 an extraordinary professor there. A year later he went as a 
visiting professor to the United States and remained there until his death in 1978 \cite{herzfeldbio}.} gave a review with the title {\em Molecular- 
and Atomic Theory of Magnetism} at the annual meeting of  German physicists in Danzig.
His comment is also due to the final remark of Ising in his publication pointing to the problems with conducting 
electrons.\footnote{Herzfeld erroneously writes Lenz as author when he cites Ising's paper, where Ising notes the problems with larger electric 
field strengths through conducting electrons.} He explains\footnote{\it Hier hat Schottky \cite{schottky} auf den zuerst von Land\'e \cite{lande}, 
Born, Heisenberg und Nordheim \cite{nordheim} in Kristall- und Molek\"ulbaufragen angenommenen Synchronismus hingewiesen.}  that the synchronism 
which is essential in Schottky's calculation was used by Land\'e \cite{lande}, Born and Heisenberg \cite{bornheisenberg}, and 
Nordheim \cite{nordheim}. Herzfeld describes the idea in the following way:\footnote{\it Wenn wir etwa zwei Wasserstoffatome haben, deren Elektronen 
in parallelen, auf die Verbindunglinie senkrechten Ebenen synchron und gleichphasig umlaufen, so wirken diese beiden infolge ihres 
Synchronismus als Dipole aufeinander ein, w\"ahrend bei ungleichen Umlaufzeiten das Dipoldglied des Potentials herausgemittelt w\"urde und nur 
ein Quadrupolglied \"ubrig bliebe...Allerdings ergibt das einfachste Modell mit synchronen Kreisbahnen in parallelen Ebenen und kubischer 
Anordnung eien solche mit abwechselnden Umlaufrichtungen, d.h. ein negatives inneres Feld.} {\it ``If we have two hydrogen atoms, for example, 
whose electrons rotate synchronously and in phase in parallel planes perpendicular to the connecting line, then these two act as dipoles on one 
another as a result of their synchronism, while with unequal rotation times the dipole member of the potential would be averaged and only 
one quadrupole member would remain. ... However, the simplest model with synchronous circular paths in parallel planes and a cubic arrangement 
results in one with alternating directions of rotation, i.e. a negative inner field."} But this conclusion of Herzfeld is based on an interaction of magnetic dipoles. 
Schottky’s argument however is based on the Coulomb interaction of the electrons whereas the nuclei constitute a fixed neutralizing background. 
This Coulomb energy is orders larger than the dipole interaction.

\section{Not only circulating but also rotating electrons}\label{VII}

\subsection{The new quantum mechanics}  

Van Vleck in his 1977 Nobel Lecture {\em Quantum Mechanics: The Key to Understand Magnetism} \cite{vleck77} said: {\it ``I would characterize this era [the 1920ties] as one of increasing disillusion and disappointment in contrast to the hopes which were so high in the years 
immediately following 1913."} And he added: {\it ``A quantum mechanics without spin and the Pauli exclusion principle would not have been enough..."}
 Indeed the ``old" Bohr-Sommerfeld quantum mechanics used three quantum integer numbers for the classical orbital motion of an electron, but this was not enough for explaining the spectrum of an atom within an external magnetic field. In order to explain experimental results in 1921 Land\'e for example introduced half-integer values for  the ``magnetic" quantum number $m$. Wolfgang Pauli who came to Lenz's group at Hamburg University in May 1922 as an assistant, was also interested in this problem. He used the relative neigbourhood to Kopenhagen and went in September 1922 for a year to Bohr's institute in Copenhagen in order to explain the anomalous Zeeman effect but he also did not succeed \cite{pauli0}. He returned in October 1923 to Hamburg where he submitted on  January 17 1924 his application for habilitation 
and on February 23 1924 he already held his inaugural lecture \cite{reich}.

So in the rest of the year 1924 Ising worked on finishing his thesis, while Pauli supervised Mensing in her thesis and developed his idea of a fourth quantum number for the electron - his {\it ``classically not describable two-valuedness in the quantum mechanical  properties of the valence electron" (`` klassisch nicht beschreibbare Zweideutigkeit der quantentheoretischen Eigenschaften des Leuchtelektrons")} \cite{pauli1} - and shortly later the exclusion principle \cite{pauli2}. 
This principle says that electrons in an atom cannot agree in all values of their quantum numbers and Pauli had to state\footnote{\it ``Eine n\"ahere Begr\"undung f\"ur diese Regel k\"onnen wir nicht geben, sie scheint sich jedoch von selbst als sehr naturgem\"a{\ss} darzubieten."}: {\it `` We cannot give a more detailed explanation for this rule, but it seems to be very natural on its own."}

Already at that time also  in connection with his findings Pauli had strong doubts on the sense of present status of quantum mechanics. 
In a letter to Bohr of December 12, 1924 he wrote: {\it ``I believe that energy and momenta values of the stationary states are something more 
real than orbits."} \cite{heilbron}. Thus one might speculate that Ising in his discussions with Pauli got more and more doubts on the validity of Schottky's idea. How fast the development was might have been seen from another  statement of Pauli  in a correspondence with Land\'e one month earlier where he describes his findings: 
{\it ``In a puzzling, non-mechanical way, the valence electron [of an alkali atom] manages to run about in two states with same k [but] with different moments."} \cite{heilbron}. 
Immediately after Pauli's publications classical interpretations of the new degree of freedom of the electron came up (for a recent discussion see \cite{giulini}). 
Pauli was very sceptical about classical interpretations of the spin by considering the electron as charged rotating sphere, which goes hand in hand with the rejection of the electron circling {{round the nucleus}} like a planet round the  {{Sun}}.

Although the spin of the electron was recognized as a two valued quantity important for the magnetic behavior of the atom (splitting of lines in the anomalous Zeeman effect and explanation of the 
Stern-Gerlach experiment) no connection to ferromagnetism or the Ising model has been made neither by Lenz nor by Pauli at that time. This had to await the formulation of the new quantum mechanics and the connection of the many body wave function to the exclusion principle. Thus at the beginning from 1925 both the microscopic basis and the macroscopic theoretical approach to ferromagnetism remained unclear. 

Even after Ising's disappointing  result that he could not prove ferromagnetism in solids by a short ranged interaction between the constituting elements of the solid it remained unclear on which level - the microscopic  or the macroscopic one - theoretical assumptions were wrong.   It was obvious that the {\it crisis} (see chapter 8 in \cite{kragh}) of the old quantum mechanics makes Schottky's arguments unacceptable. And indeed within the following years a complete new quantum mechanics was developed. There is a big historic literature on these developments (see e.g. the contributions to \cite{hoddeson} especially those concerned with `The Development of the Quantum Mechanical Electron Theory of Metals 1926 - 1933' by L. Hoddeson, G. Baym, and M. Eckert, `Magnetism and Magnetic Materials' by St. T. Keith and P. Qedec, and `Collective Phenomena' by L. Hoddeson, H. Schubert, St. J. Heims, and G. Baym)). In the following only those ideas are presented which are important for replacing Schottky's argumentation.

Heisenberg notes three problems in quantum mechanics, which were connected and caused  great difficulties to solve them: {\it ``the anomalous Zeeman effect, the exclusion principle and the dualism between particles and waves."} 
On June 9th  1925 Heisenberg wrote from G\"ottingen to Pauli\footnote{\it ``Es ist wirklich meine \"Uberzeug[un]g, dass eine Interpretation der Rydberg Formeln in 
Linien von Kreis u. Ellipsenbahnen in klassischer Geometrie nicht den geringsten physikalischen Sinn hat und meine ganzen k\"ummerlichen Rechnungen gehen dahin, 
den Begriff Bahnen, die man doch nicht beobachten kann, restlos umzubringen und geeignet zu ersetzen."} in Hamburg: {\it ``It is really my conviction that an 
interpretation of the Rydberg formulas in lines of circles and elliptical orbits in classical geometry does not have the slightest physical sense and all 
my poor calculations go to the point to kill completely the term orbits that cannot be observed anyway, and to replace it properly."} 
Pauli's answer is lost. Heisenberg submitted on July 29 1925 his paper \cite{heisenberg1} and the main point was the avoiding of non-observable quantities 
in the formulation of the theory. Thus quantities like orbits and orbital periods are replaced by matrix-like schemata (the mathematical formulation by 
matrices followed later by Born, Jordan, and Heisenberg) of the observed radiation of the electron between the stationary states (for further explanations 
see \cite{aitchison}).

Erwin Schr\"odinger developed alternatively his wave mechanics \cite{schroedinger1926} and showed its equivalence to Heisenberg's version.  The modulus square of the wave function gives the probability density of finding the electron in a certain position in space. This allows a  visualization of the distribution of the electric charges in the atoms (see later Fig. \ref{psih2}) different from the old orbits.

\subsection{Heisenberg's exchange interaction}

At the end of Herzfeld's review on magnetism \cite{herzfeld} he remarks after commenting on Schottky's paper:\footnote{\it So ist die Frage nach der Natur des inneren 
Feldes in ferromagnetischen K\"orpern noch offen.} {\it ``So the question of the nature of the internal field in ferromagnetic bodies is still open."} 
In order to settle this question J. G. Dorfman \cite{dorfman} started in 1927  experiments to prove that 
the molecular field in ferromagnets could not be of magnetic origin. He wrote: {\it ``From many points of view it seems evident that the so-called 'molecular field' 
introduced by P. Weiss into the theory of ferromagnetism cannot be purely magnetic. ... As an instrument for studying the intrinsic fields,
a narrow beam of real free electrons, $\beta$-particles, was chosen. If a magnetic field exists inside a magnetized foil of nickel, it is evident 
that the beam of $\beta$-particles passing through the foil will be deflected. .. Thus it can be claimed that no magnetic field exceeding 105 gauss 
exists in a ferromagnetic substance. Further experiments on the passage and scattering of $\beta$-particles in ferromagnetic substances are in
progress.''} Kapitza in the obituary for Dorfman \cite{kapitza} judged: {\it ``This experimentum crucis implied that the molecular field can be of only electrical nature, as Frenkel [citation of \cite{frenkel}] and W. Heisenberg became the first to demonstrate (a year later\footnote{We corrected the year given in the citation from 1938 to 1928} in 1928)."} Surprisingly no reference was made to Schottky's paper although he had indicated a way out of this dilemma.

\begin{figure}[h]\centering
\includegraphics[height=0.30\textwidth]{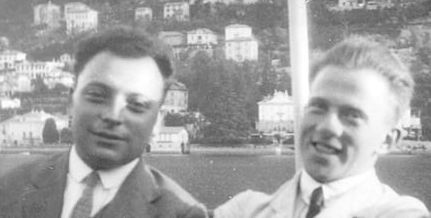}
\caption{\label{ph27} Pauli and Heisenberg 1927 \copyright Cern Geneva (cutted).}
\end{figure}

The problem of the microscopic basis of the interaction of the elementary magnetons in ferromagnets remained although it was now shifted from the orbit of the electron to the rotating electron itself.  {\it ``The origin of Heisenberg's interest in ferromagnetism is uncertain. One source is likely his friendship with physicists in Hamburg, for whom magnetism 
was a favorite topic"} (p.305 in \cite{hoddeson87}). Especially the contact to Pauli (Fig. \ref{ph27}) kept him informed on Ising's thesis. Beginning on 3 May 
1928 Heisenberg sent a series of letters\footnote{Archive on the web \url{https://archiv.heisenberg-gesellschaft.de/}} 
concerning his present work on ferromagnetism. This is an example how Heisenberg used Pauli as a referee for his paper:\footnote{\it ``Ich habe nie 
eine Arbeit ver\"offentlicht, ohne dass Pauli sie vorher gelesen hatte. Sagte er, sie sei falsch, so konnte sie immer noch richtig sein; aber 
dann war Vorsicht am Platze."} {\it ``I never have published a work without Pauli reading it first. If he said it was wrong, it could still be right; 
but then caution was needed"} \cite{PBW1}. Heisenberg sent his paper for publication on  May 28 1928 \cite{heisenberg28}.

However already on November 4 1926 he had communicated some ideas about ferromagnetism to Pauli. He wrote\footnote{\it Die Idee ist die: Um die Langevinsche Theorie des Ferromagn. zu brauchen, muss man eine grosse Kopplungskraft zwischen den spinnenden Elektronen annehmen (es drehen ja \underline{nur} \underline{diese} sich). Diese Kraft soll wie beim Helium von der Resonanz indirekt geliefert werden. Ich glaube, man kann allgemein beweisen: Parallelstellung der Spin-vektoren gibt stets \underline{kleinste} Energie. Die Energieunterschiede, die in Betracht kommen, sind von \underline{elektrischer} Gr\"o{\ss}enordnung, nehmen aber mit zunehmenden Abst\"anden \underline{sehr} rasch ab. Ich hab das Gef\"uhl, (ohne das Material auch nur entfernt zu kennen) dass dies im Prinzip zu einer Deutung des Ferromagnetismus ausreichen k\"onnte.} (underlining due to the original): {\it ``The idea is this: In oder to use Langevin's theory of ferromagn., one has to assume a strong coupling force between the spinning electrons (\underline{only} \underline{these} turn around). As with helium, this force should be supplied indirectly by the resonance. I think one can generally prove: parallel positioning of the spin vectors always gives \underline{smallest} energy. The energy differences that come into consideration are of the \underline{electrical} order of magnitude, but decrease \underline{very} rapidly with increasing distances. I have the feeling (without even knowing the material remotely) that this could in principle be sufficient for an interpretation of ferromagnetism."} One may remember the following: Curie studied ferromagnetism and its dependence on an external magnetic field and temperature. Based on this Langevin formulated 1905 even before Bohr's atomic model a microscopic theory of elementary interacting magnetons and treated them by statistical theory. Weiss made then the important step of introducing an internal field by the elementary magnetons.

\begin{figure}[htbp]\centering
\includegraphics[height=0.25\textwidth]{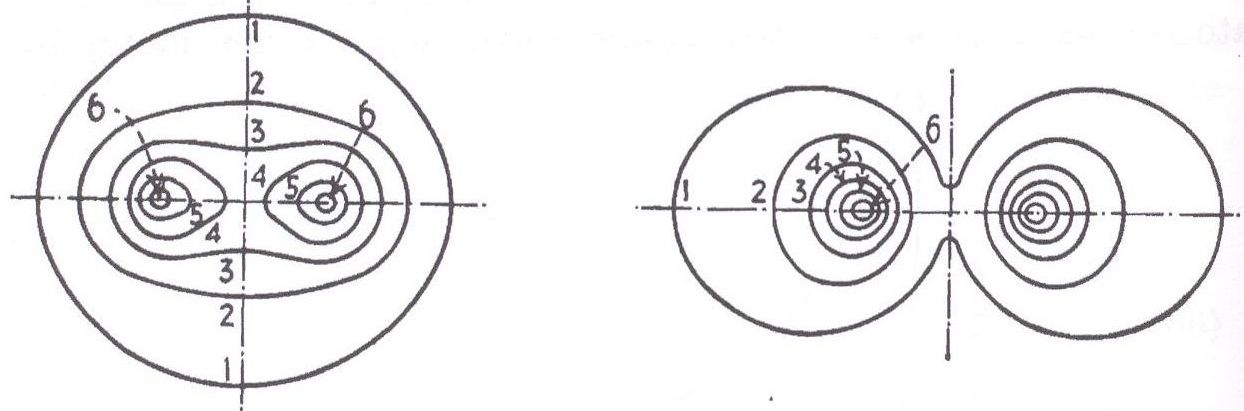}
\caption{\label{psih2} The wave function of the hydrogen molecule for antiparallel (left) and parallel (right) spin orientation. From the charge distribution it is seen that the Coulomb energy for the parallel case is lower. (From \cite{bitter} Fig. 67 p. 134).}
\end{figure}

It is surprising that no reference to Schottky's earlier idea is made at that place of speculation. Both of them were well informed about Schottky's paper (see sect. \ref{V}). 
Anyway more than one and a half years later the above mentioned idea was elaborated and published by Heisenberg without citing Schottky's paper.
In our opinion, this shows not the negligence of priority but rather a fundamental change of the physical concepts made by the new quantum mechanics.
Heisenberg based his calculations on the method developed 1927 by Heitler and London. Of essential importance 
in these calculations is the Pauli principle and the symmetry property of the whole many body eigenfunction. Fig. \ref{psih2} taken from Bitter's 
{\em Introduction to Ferromagnetism} of the year 1937 shows the difference of these functions for the two possible spin states - parallel or antiparallel - 
from which the difference in the charge distribution is obvious.\footnote{See also the 
animated graphics shown by Vadym Zayets, especially Fig. 6 there: Vadym Zayets {\em Exchange Interaction} webpage: 
\url{https://staff.aist.go.jp/v.zayets/spin3_47_exchange.html}
}  One important conclusion of Heisenberg's calculation was that ferromagnetism is then only possible for  lattice types for which an atom has at least 
eight neighbors and he then notes {{Fe, Co, Ni as examples.}}

\begin{table}[h]\centering
\begin{tabular}{|c|c|c|c|c|}\hline
\multirow{2}{20mm}{\sc Schottky}&\multirow{2}{20mm}{ orbital } & orientation of & \multirow{2}{22mm}{synchronization}&\multirow{2}{18mm}{Coulomb energy}\\ 
 && circulation &  &\\ \hline\hline
\multirow{2}{23mm}{\sc Heisenberg}&{wave function}&\multirow{2}{20mm}{spin part}  & \multirow{2}{22mm}{Pauli principle} & \multirow{2}{15mm}{Coulomb energy}\\ 
&spacial part &&&\\ \hline
\end{tabular}
\caption{\label{comp}The comparison of Schottky's idea with the new quantum mechanics.}
\end{table}

Heisenberg uncovered the microscopic `force' which was thought for over the years but needed a complete change in the underlying microscopic theory, the quantum mechanics. Classical concepts were no longer sufficient to explain the observed behavior. The common concept of Schottky's and Heisenberg's explanation was the principle that the `system' chooses the state of lowest energy that is the lowest value of the Coulomb energy. How this could come about was postulated in Schottky's idea in the synchronization of the two electrons and the direction of the circulation on their orbits. In Heisenberg's concept no orbits exist but static charge distributions depending on the spinning electron. The circulation of the electron which produced the elementary magneton is replaced by the spinning electron and its resulting magnetic moment. The synchronization is replaced by the Pauli principle and the genuine symmetry property of the wave function of the electron. Table \ref{comp} summarizes the correspondence of both ideas.

In the contribution \cite{bopp} {\em Fifty Years of Quantum Theory} Friedrich Arnold Bopp remarks commenting Heisenberg's {\it exchange force}: {\it ``Whereas the analogous question concerning phase relations between rotating electrons [see Sect. \ref{V}] could not be solved with the older quantum theory, wave mechanics offered certain correlations, that is, the symmetry or antisymmetry in the position coordinates of the wave function.''} And somewhat later he says: {\it ''All these investigations are based on the Pauli principle''}
This rejection of the idea of synchronism, its replacement by the Pauli principle and the spin of the electron had changed physics of the microscopic basis 
of the Ising model but did not change the statistical mechanics of the the model. However it allowed a new mathematical formulation.

\section{Solvay 1930 and Pauli's talk}\label{VIII}
About two months after submitting his paper on ferromagnetism Heisenberg on July 31, 1928 came back to `ferromagnetism' because he had written a Festschrift for Sommerfeld on this topic \cite{heisenbergfest}. He was unsatisfied with the Gaussian approximation he had made. {{He called it in a letter to Pauli {\it`a very unpleasant fraud' (ein sehr unerfreulicher Schwindel)}.}} Then he came to the comparison of his model with Ising's model. 
He concluded erroneously as we know now:\footnote{\it ``das hier zugrunde gelegte Modell weist tats\"achlich gro{\ss}e \"Ahnlichkeit mit dem Isingschen Modell 
(Wechselwirkung nur zwischen Nachbaratomen) aus und unterscheidet sich von jenem im wesentlichen {\em nur durch den Wert $z$}, d.h. durch die Anzahl der Nachbarn, 
welche ein Atom umgeben."} {\it ``the  model presented here is actually very similar to Ising's model (interaction only between neighboring atoms) and differs from that model essentially 
{\em only in the value $z$}, i.e. in the number of neighbors surrounding an atom "} [\cite{heisenbergfest}  pp. 121 - 122,] accentuation by the authors]. It is not the value of $z$, which is the pivotal question but the type of the order parameter. It is a scalar in the case of the Ising model and a vector in the case of the Heisenberg model. The two models belong to different universality classes as we know today and show different critical behavior at their phase transitions in $d=3$. E.g. the specific heat  diverges for the Ising model whereas it is finite for the Heisenberg model. In the letter Heisenberg added further criticism regarding Ising's paper. 
In fact Ising describes his three dimensional model as a accumulation of planes which itself are accumulations of chains. In order to proceed with the calculations he made approximations. These in fact are chosen in such a way that effectively he only calculates a noninteracting accumulation of chains, which lead to the same result a single chain multiplied by the number of chains. No information on these approximations used was given in Ising's  publication.

On  January 2 1928 Dirac had submitted his paper {\em The Quantum Theory of the Electron} \cite{dirac1928} where he {\it ``would like to find some incompleteness in the previous methods of applying quantum mechanics to the point-charge electron such that, when removed, the whole of the duplexity phenomena [the spin of the electron] follow without arbitrary assumptions."}  
Dirac explained in an interview with Hund \cite{hundDirac} how he had to introduce the spin of the electron into the desired complete equation for the electron. Dirac had constructed a transformation theory containing the first time derivative instead of the second one as in the Schr\"odinger Klein-Gordon equation. Dirac explained (for a more detailed explanation of these ideas see chapter 3 in \cite{kragh2}): 
{\it ``that forced me into a different kind of equation and this different equation brings in the spin of the electron. It was very unexpected to me to see the spin appearing in that way. [ spin means both moment of momentum and magnetic moment Hund adds] I thought one would have a satisfactory theory without spin and then one would proceed to a more complicated theory."}
Thus Dirac completed the theoretical concepts necessary to understand the microscopic basis  of ferromagnetism. 

{\it ``With the exception of the great mathematicians' congress in Paris (1900), the Solvay conferences (founded 1911) were the first truly international conferences."} \cite{schirrmacher} (see also \cite{marage}).  They were located in Brussels and planned every third year. After the Great War they started again 1921. The 5th Solvay conference in 1927 dealt with the foundations of the physical world on microscopic scale - the foundations of the new quantum mechanics - and was the last under the presidency of H. A. Lorentz. The discussion proceeded on the 6th Solvay conference under the presidency of P. Langevin but the theme shifted to the microscopic basis of the macroscopic phenomenon {\sc magnetism}.  In fact Elliott \cite{elliott} counted the 6th Solvay conference as the zeroth International Conference on Magnetism before the conference in Strasbourg 1939. 


Pauli was invited to give a review talk \cite{pauli1930} on the 6th Solvay conference in Brussels from October 20 to 25 with the title {\em The Quantum Theories of Magnetism. The Magnetic Electron (Les Th\'eories Quantiques du Magn\'etisme. L'\'electron Magn\'etique.)}. 
He starts his introduction with the following statement:\footnote{\it Un r\'esultat essentiel des nouvelles th\'eories du magn\'etisme dans
ce domaine est d’avoir pu fonder sur la d\'ecouverte du moment
propre de l’\'electron (moment de pivotement ou spin) une explication g\'en\'erale de deux ph\'enomènes magn\'etiques caract\'eristiques de l’\'etat solide.} {\it ``An essential result of the new theories of magnetism in
this field is to have been able to base on the discovery of the momentum
characteristic of the electron (pivoting momentum or spin) a general explanation of two magnetic phenomena characteristic of the solid state."} Then he points to the 
papers of Heisenberg and concludes:\footnote{\it ``Les th\'eories ant\'erieures ne permettaient pas de pr\'evoir si un corps solide donn\'e devait être ferro, para ou diamagn\'etique, ni de
calculer l’intensit\'e de son aimantation."} {\it ``Previous theories did not make it possible to predict whether a given solid body should be ferro, para or diamagnetic, 
nor to calculate the intensity of its magnetization."} 

He cites a paper of Felix Bloch \cite{bloch1930} and gives a result for the magnetization. Important is the remark concerning the dimensionality of the system:\footnote{\it ``Le r\'esultat essentiel de cette 
th\'eorie est que la tridimensionalit\'e du r\'eseau est n\'ecessaire pour l’apparition du ferromagn\'etisme et que, dans ce cas, il est suffisant que 
l’int\'egrale de permutation soit positive."} {\it ``The essential result of this theory is that the three-dimensionality 
of the lattice is necessary for the appearance of ferromagnetism and that, in this case, it is sufficient that the permutation integral is positive.}" 
Indeed  the Mermin-Wagner theorem \cite{merminwagner}, proved 36 years later, corroborated this expectation.

Then he compares with the linear chain model of Ising and says [p.210 in \cite{pauli1930}]: {\it ``We can say that the energy, or Hamilton's function is in this case"} (in fact he means the Heisenberg model)
\begin{equation}
H=-A\sum_k(\sigma_k,\sigma_{k+1}) \, , \label{imodel}
\end{equation}
{\it ``where $\sigma_k$ represents the pivot vector of the $k${th} electron"} and $A$ is given by the exchange integral. The meaning of the spin vector matrix $\sigma_k$  is explained in the chapter on the Dirac equation, the relativistic quantum mechanical equation for the electron (see p. 228). The bracket means the scalar product of the spin vectors $\sigma_k$ and $\sigma_{k+1}$.

He then points to the difference of Heisenberg's model\footnote{\it ``Dans le calcul d’Ising, d\'evelopp\'e au point de vue de l’ancienne th\'eorie des quanta, les composantes des perpendiculaires à la direction du champ sont consid\'er\'ees comme nulles, tandis que dans la nouvelle m\'ecanique cette composante n’est pas commutable avec celle qui correspond à la direction du champ. Malgr\'e cette diff\'erence, il est très vraisemblable qu’une
extension de la th\'eorie d’Ising au cas d’un r\'eseau à trois dimensions donnerait du ferromagn\'etisme même au point de vue classique."}: {\it ``In the calculation of Ising, which are developed from the point of view of the old quantum theory, the components perpendicular to the direction of the field are considered zero, while in the new [quantum] mechanics this component does not commute with that corresponding to the direction of the field. Despite this difference, it is very likely that an extension of Ising's theory to the case of a three-dimensional network would give ferromagnetism even from the classical point of view."} 

Thus in the formulation of the Ising model given in Eq. (\ref{imodel}) 
only the product of the two vector components in the external field direction is taken. If one  {{chooses}} the representation of the third or $z$-component of the spin vector one has
\begin{equation} 
\sigma_k= \left( \begin{array}{rr}
1 & 0 \\ 
0 & -1  \\
\end{array}\right)  \, .
\end{equation}
Pauli had introduced the matrix representation of the spin components already 1927 in order to include the spin also in Schr\"odinger's version of the quantum mechanics 
\cite{pauli1927}. Now in the Hamiltonian given in Eq. (\ref{imodel}) the Ising model has already found its final mathematical representation.

\begin{figure}[ht]\centering
 \includegraphics[height=0.5\textwidth]{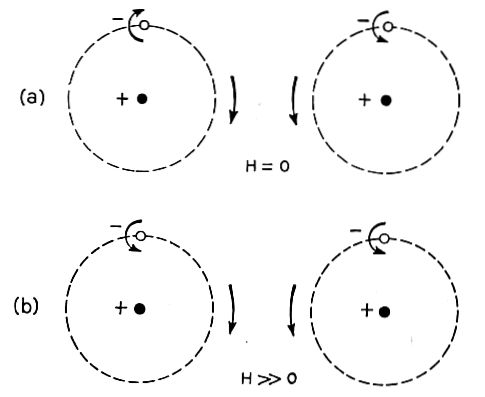}
\caption{\label{bozoroth} ``In the common ferromagnetic materials a change in magnetization is effected by a change in the direction of electronic spin, not in the direction of motion of the electron in 
its orbit" \cite{bozorth}.}
 \end{figure}

Although the mathematics had  changed a lot for computing the presumed ferromagnetic phase by statistical mechanics the old classical descriptive pictures remained. An example for that can be found in the review article \cite{bozorth} by Bozorth  in 1940 (see Fig. \ref{bozoroth}). 
One should add to the description of Fig. \ref{bozoroth} that there is a force between the spinning electrons which is related to the symmetry property of the whole eigenfunction for the two electrons induced by the Pauli principle. Unfortunately it is difficult if not impossible to make this visible in an image.
 
 One year later a crucial step was made by Kramers and Wannier \cite{kramerswannier} by introducing the transfer matrix  and  the duality symmetry of the Ising model to prove that there is indeed a phase transition. It is astonishing that with the duality it can be shown (see \cite{mattis} p. 236) that Ising's chain in zero magnetic field corresponds to Schottky's two level system \cite{schottky}, which could have been another reason to cite Schottky's paper when the specific heat of the chain is calculated.
 
In order to see this correspondence one introduces new spin variables called link-spins
 \[\mu_k=\sigma_k\sigma_{k+1} \,.\]  
The Hamiltonian (\ref{imodel}) {then reads}
\[H=-A\sum_k \mu_k \, , \]
{it corresponds to a system of non-interacting spins} which can be in two different energy states - the two-level system considered by Schottky \cite{schottky}.

We know since the paper of Kramers and Wannier that one may look for a phase transition by calculating physical quantities different from the magnetization, like the specific heat or the correlation function. This allows to {analyse}  phase transitions without introducing an external field.
 
\section{Conclusion}\label{IX}

We have presented Schottky's idea about a short ranged interaction of elementary magnetons caused by circulating atomic electrons within the old quantum mechanics. This interaction prefers parallel magneton arrangement due to minimizing the Coulomb energy. This idea was outperformed by the new development of quantum mechanics. The discovery of  electron spin as a property of elementary particles,  the new description of electron states by wave functions and their interpretation led a new formulation of this interaction by Heisenberg. Unfortunately the success of new quantum mechanics blurred the traces of the old ideas. We tried to recover them.

This also led to a new formulation of the Ising model given by Pauli at the Solvay conference and subsequently to new mathematical 
consideration within the statistical mechanics (e.g. via the transfer matrix). However although the short range interaction was clarified, the emergence of a long range order in the macroscopic systems remained, but now as a problem of statistical physics. Nevertheless it turned out that the understanding of critical phenomena again had to come back to the concepts of elementary particle physics in  the development of  renormalization group theory.

\section*{Acknowledgment}
R.F. thanks Sigismund Kobe for discussions on the Schottky anomaly.
The authors are thankful to Bertrand Berche and Ralph Kenna for a longstanding
collaboration on a history of Ising model and for reading and discussing this
paper prior to publication. We also highly acknowledge the support by the Austrian 
Agency for International Cooperation in Education and Research (OeAD) and the Ministry
of Education and Science of Ukraine via bilateral Austro-Ukrainian grant number 
UA 09/2020. Yu.H. acknowledges support
of the JESH mobility program of the Austrian Academy of Sciences and hospitality
of the Complexity Science Hub Vienna when finalizing this paper.

\section*{Authors contributions}
All the authors were involved in the preparation of the manuscript.
All the authors have read and approved the final manuscript.

\section*{Appendices}

\begin{appendix}
\section{Timeline}\label{timetable}
In the table below we give the timetable of some of the events discussed in this paper.

\begin{enumerate}
\item[1920] September 19 - 25 Lenz presents his paper in Nauheim at the 86th meeting of the Society of German Natural Scientists and Physicians published  December 5 \cite{lenz} 
\item[1921] August Stern presented the experimental idea to prove spatial quantization
\item[1921] Lenz gets chair at the University Hamburg
\item[1921] April 21  Ising registered  at the University Hamburg
\item[1922] February 7 - 8 Stern-Gerlach experiment performed,  publication received on March 1 \cite{gerlach}.
\item[1922] Schottky is until 1923 a regular assistant at Physikalisches Staatslaboratorium Hamburg
\item[1922] September 21 Schottky's paper presented in Leipzig at the 87th meeting of the Society of German Natural Scientists and Physicians \cite{schottky}
\item[1922] Ising starts his thesis
\item[1922] May Pauli employed at the University Hamburg
\item[1922] September Pauli starts one year studies in Copenhagen, Ising replaced duties of Pauli
\item[1923] October Pauli back in Hamburg
\item[1924] February 23 Pauli holds his inaugural lecture {\em Quantum Theory and Periodic System of the Elements}
\item[1924] July 28 Ising calls for doctorate (thesis completed)
\item[1924]  December 8 Ising submits his publication \cite{Ising25}
\item[1925] Ising has left Hamburg and accepted a position at the General Electric Company in Berlin
\item[1925] January 15 Pauli submitted his publication \cite{pauli1,pauli2}
\item[1925] July 29 Heisenberg submitted his paper on the new quantum mechanics (the re-interpretation of the old one) \cite{heisenberg1}
\item[1925] September 10 - 16 Herzfeld comments Schottky's and Ising's paper at the Third Deutsche Physikertagung   in Danzig \cite{herzfeld}
{ \item[1926] January 27 Schr\"odinger submitted his first paper on quantum mechanics \cite{schroedinger1926}}
\item[1926] January Pauli became professor at the University Hamburg.
\item[1926] November 4 Heisenberg sends the letter to Pauli with the idea concerning the coupling force of the electrons spin
\item[1927] January 6 Dorfman submitted his paper published on March 5 \cite{dorfman} 
\item[1927] May 3 Pauli submitted his paper where he introduced the spin matrices \cite{pauli1927}
\item[1928] January 2 Dirac's paper on the equation for the electron \cite{dirac1928} was received 
\item[1928] April 1 Pauli gets the chair at ETH Zurich (accepted by Pauli February 2.) 
\item[1928] May 3 Start of a series of letters from Heisenberg  to Pauli  discussing Heisenberg's research on ferromagnetism 
\item[1928] May 20 Heisenberg submits his paper on ferromagnetism \cite{heisenberg28} 
\item[1928] May 31 Heisenberg writes a letter to Pauli concerning Ising's model 
\item[1930] October 20 - 25 Pauli presents the Hamiltonian of the Ising model at the Solvay Conference in Brussels  \cite{pauli1930}
\end{enumerate}

\section{Calculation for a general configuration of two electron orbits \label{calculation}}

\begin{figure}[h]\centering
\includegraphics[height=5.5cm]{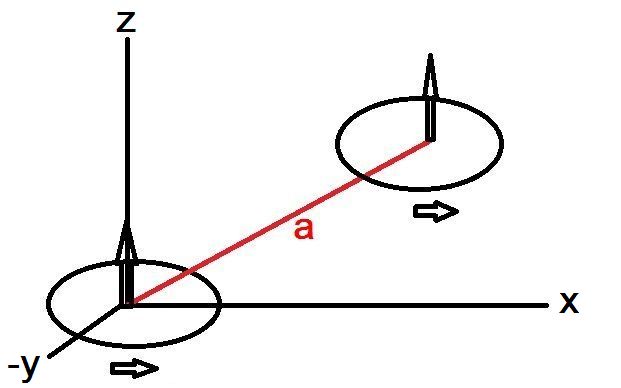}
\caption{
Two atoms in the $xz$-plane at distance a (red line) with electrons circling in the same direction. Then the produced elementary magnetons are parallel to each other. If the electron circling around the origin changes its rotation direction then the two elementary magnetos are antiparallel. Schottky showed that then the Coulomb energy is larger compared to the case of parallel elementary magnetons. \label{alpha}}
\end{figure}

We calculate the distance between two electrons rotating with frequency $\omega$ on two circles with radius $r$ whose centers are at distance $a$ (see Fig. \ref{alpha}) and the spacial angle $\alpha$.  The phase between the two rotations is fixed to $\phi$. The coordinates read
 \begin{align}
x_1&= r \cos\omega t & x_2&= r \cos(\omega t+\phi)+a\cos\alpha\\
y_1&=r \sin\omega t  & y _2&=r \sin(\omega t+\phi)\\
z_1&=0 & z_2&=a \sin\alpha \, .
\end{align}
Changing the notations: $\omega t=\tau$, $R=r/a$  we get for the distance if both electrons rotate in the same direction:
\begin{equation}
d_1(\tau,\phi,\alpha)=\sqrt{\big(R(\cos(\tau +\phi)-\cos\tau)+\cos\alpha\big)^2+R(\sin(\tau +\phi)-\sin\tau)^2+\sin^2\alpha }'\, ,
\end{equation}
where the distance now is measured in the length of $a$.
Changing the direction of  circulation of one electron  [$\tau \to -\tau$] (e.g. electron 1) one gets for the distance
\begin{equation}
d_2(\tau,\phi)=\sqrt{\big(R(\cos(\tau +\phi)-\cos\tau)+a\cos\alpha\big)^2+R(\sin(\tau +\phi)+\sin\tau)^2+\sin^2\alpha}\, .
\end{equation}
Using 
\[\cos(\tau +\phi)-\cos\tau=-2\sin(\phi/2)\sin(\tau+\phi/2)\, ,\]
\[\sin(\tau +\phi)-\sin\tau=2\sin(\phi/2)\cos(\tau+\phi/2)\, ,\]
\[\sin(\tau+\phi)+\sin\tau = 2\sin(\tau+\phi/2)\cos(\phi/2)\, , \]
one arrives at
\begin{equation}
\small{d_1(\tau,\phi,\alpha)=\sqrt{\big(R(-2\sin(\phi/2)\sin(\tau+\phi/2)+\cos\alpha\big)^2+4R^2\sin^2(\phi/2)\cos^2(\tau+\phi/2)+\sin^2\alpha }}
\end{equation}
and
\begin{equation}
\small{d_2(\tau,\phi)=\sqrt{\big(R(-2\sin(\phi/2)\sin(\tau+\phi/2)+\cos\alpha\big)^2+4R\sin^2(\tau+\phi/2)\cos^2(\phi/2)+\sin^2\alpha}\, .}
\end{equation}
Further reducing these expressions leads to
\begin{eqnarray}
d_{1}(\tau,\phi,\alpha)=\sqrt{1+4R^2\sin^2(\phi/2)-4R\sin(\phi/2)\sin(\tau+\phi/2)\cos\alpha}\, , \\
d_{2}(\tau,\phi,\alpha)=\sqrt{1+4R^2\sin^2(\tau+\phi/2)-4R\sin(\phi/2)\sin(\tau+\phi/2)\cos\alpha} \, .
\end{eqnarray}
If one averages the two distances with respect to $\tau$ over one period by simply taking the mean of $\sin(x)$ 
and $\sin^2(x)$ the space angle $\alpha$ drops out
\begin{eqnarray}\label{d1}
\bar{d}_{1}=\sqrt{1+4R^2\sin^2(\phi/2)}\, ,\\ \label{d2}
\bar{d}_{2}=\sqrt{1+2R^2}\, .
\end{eqnarray}
The distance $\bar{d}_{1}$ between the electrons when they rotate in the same direction depends on the phase and is largest for the phase $\phi=\pi/2$.
Taking now distances (\ref{d1}), (\ref{d2})  for calculating the Coulomb energy leads to the result of Eq. \eqref{ourresult}.
For $R=1/4$ [$R=1/2$] this is of the order of
\begin{equation}
\Delta E=0.05[0.11]\frac{e^2}{a}\, .
\end{equation}
Although this difference is smaller than found by Schottky. His values are in the range of 
\begin{equation}
\Delta E=0.06[0.3]\frac{e^2}{a}\, ,
\end{equation}
 but both estimates lead to Curie temperature which remains much larger than obtained from the magnetic dipole interaction. 
 It also has the advantage that it does not depend on the configuration of the orbits in a cubic crystal structure.

Schottky \cite{schottky} claimed:\footnote{\it Bei Gleichl\"aufigkeit und Gegenl\"aufigkeit der beiden Bahnen gibt es je eine Phase, die der kleinsten elektrostatischen 
Wechselwirkungsenergie der beiden Elektronen entspricht.} 
{\it ``If the two orbits are in the same direction and in opposite directions, there is in each case a phase corresponding to the smallest electrostatic 
interaction energy of the two electrons."}

It remains open for which distance he calculated this time independent energy. One possibility is to calculate this energy for the mean 
distance. Another possibility could be to take the  energy of the smallest distance during a circulation and to choose the phase in such a 
way that this smallest distance is maximised. So to say the possible smallest hill the electrons have to overcome during circulation. 
This would lead for the configuration (a) to Schottky's result but with $k$ equal $2$. 

\begin{figure}\centering
\includegraphics[height=5.5cm]{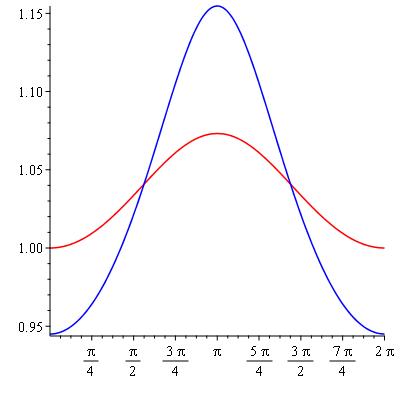}\hspace{0.1cm}\includegraphics[height=5.5cm]{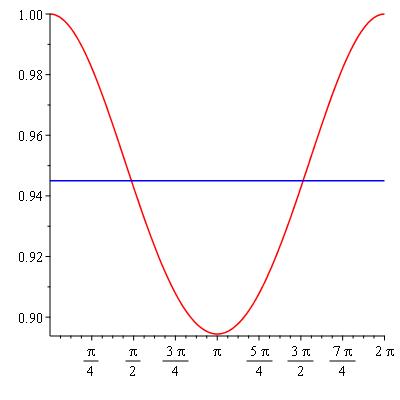}
\caption{The scaled mean of the Coulomb energy for two atoms beside  and above each other for different phases (left and right pictures, correspondingly). Red curves for electrons circling in the same direction (parallel magnetic moments), blue curves for  electrons circling in different directions (antiparallel magnetic moments). 
The values chosen are for $R=0.25$ in both cases and  $\alpha=0$ {{(left)}} and  $\alpha=\pi$ {{(right)}}. \label{besideabove}}
\end{figure}

One may improve the above calculations by averaging the Coulomb energies (the inverse distances) rather than just the distances. 
This leads to calculating elliptic integrals and numerical calculations are necessary. This calculation has been performed and the result for 
two configurations is shown in Fig. \ref{besideabove}.
Two regions of synchronizing are seen in the figure. Around the phase $\phi=\pi$ the parallel arrangement of the elementary magnetic moments has the lowest energy; 
around the phase $\phi=0$ the antiparallel arrangement has the lowest energy. Thus, if taken serious, the model contains already a possibility for 
ferromagnetic and antiferromagnetic ordering.  However antiferromagnetism was found only later in the 30's by Louis Neel. This was also recognized \cite{serchinger} by Schottky, who named this type of ordering {\it neutromagnetic (neutromagnetisch).}

Concerning the quantum mechanical background of his idea Schottky writes:\footnote{\it Die genauere modellm\"a{\ss}ige Untersuchung aller dieser Fragen wird 
nicht nur zu schwierigen Rechnungen f\"uhren, sondern auch die Kenntnis von allgemeinen Bewegungs- und Quantengesetzen erfordern \"uber die wir z. Zt. 
noch in keiner Weise gen\"ugend unterrichtet sind.} {\it ``The more exact model-based investigation of all these questions will not only lead to difficult 
calculations, but also require knowledge of general laws of motion and quantum, about which we are currently in no way sufficiently informed.''}

\end{appendix}

%

%
%
%
%

\end{document}